\documentstyle[11pt]{article}
\begin{document}
\vskip .3cm
\centerline{\LARGE Self-dual variables, positive semi-definite action,}
\vskip .3cm
\centerline{\LARGE and discrete transformations}
\vskip .3cm
\centerline{\LARGE in four-dimensional quantum gravity}
\vskip .3cm
\rm
\vskip.2cm
\centerline{Chopin Soo ${}^\dagger$}
\vskip .3cm
 \centerline{\it  Center for Gravitational Physics and Geometry}
\centerline{\it Department of Physics}
 \centerline {\it Pennsylvania State University}
\centerline{\it University Park, PA 16802-6300, USA}
\vskip .3cm
\vskip .3cm
\vskip .3cm
\vskip .3cm
\vskip .3cm
\vskip .3cm

A positive semi-definite Euclidean action for arbitrary four-topologies can
be constructed by adding appropriate Yang-Mills and topological terms to
the Samuel-Jacobson-Smolin action of gravity with (anti)self-dual variables.
Moreover, on-shell, the (anti)self-dual sector of the new theory
corresponds precisely to all Einstein manifolds in four dimensions.
The Lorentzian signature action, and its analytic continuations are also
considered. A self-contained discussion is given on the effects of discrete
transformations C, P and T on the Samuel-Jacobson-Smolin action, and
other proposed actions which utilize self- or anti-self-dual variables
as fundamental variables in the description of four-dimensional gravity.

\vskip .3cm
\vskip .3cm
PACS number(s): 04.60.-m, 04.20.Cv, 11.15.-q, 11.30.Er
\vskip .3cm
(To appear in Phys. Rev {\bf D}).
\vfill
${}^\dagger$Electronic address: soo@phys.psu.edu
\eject

\section*{I. Introduction}

Self- or anti-self-dual variables which exploit the unique properties of
four dimensions have been proposed as fundamental variables in the description
of classical and quantum gravity \cite{js, sam, aa}. Not long
after the simplification of the constraints of general relativity achieved
with these variables was announced \cite{aa}, the covariant
action was found by Jacobson and Smolin \cite{js}, and by Samuel \cite{sam}.
This first order action with independent left-handed (primed or dotted)
connection and veirbein as fundamental variables reproduces
the Ashtekar variables and constraints \cite{aa} naturally.
The resultant equations of motion are the same as the Einstein
field equations in four dimensions.

Furthermore, all Einstein manifolds in four dimensions are described by
Ashtekar connections which are (anti)self-dual with respect to the metrics of
the solutions. This is somewhat surprising since not all Einstein manifolds
have Riemann-Christofel curvature tensors which are (anti)self-dual.
In four dimensions, the four-index (antisymmetric in pairs)
Riemann-Christofel curvature tensor, $R_{\alpha\beta\mu\nu}$, can be
dualized on the left and on
the right;\footnote{These can be thought of as internal and external
duality transformations if we consider the Riemann curvature two-form
$R_{AB\mu\nu}dx^\mu\wedge dx^\nu$ instead of $R_{\alpha\beta\mu\nu}$.}
and it can be viewed as a $6\times6$ matrix mapping $\Lambda^{\pm}_2
\rightarrow
\Lambda^{\pm}_2$ of the $\pm$ eigenspaces, $\Lambda^{\pm}_2$, of the Hodge
duality operator \cite{as}
\begin{equation}
\left(\matrix{F^+ &C^+\cr C^- &F^-}\right)
\end{equation}
Here $F^+(F^-)$ is self-dual (anti-self-dual) with
respect to both left and right duality transformations while $C^+(C^-)$
is self-dual (anti-self-dual) under left duality and
anti-self-dual(self-dual) under right duality transformations.
A metric is Einstein iff the matrix is block diagonal i.e. iff the
$3\times3$ blocks $C^\pm$ vanish \cite{as}.
On-shell, $F^-$, which is {\it doubly} (anti)self-dual, is precisely the
curvature of the Ashtekar connection \cite{css}. Thus, {\it all} Einstein
manifolds are described by Ashtekar connections which are (anti)self-dual.

This raises the interesting question of whether it is possible to
construct a Yang-Mills-like theory based on a left-handed
(or (anti)self-dual) connection with the property that, on-shell, the
(anti)self-dual sector of the theory corresponds precisely to
{\it all} Einstein manifolds in four dimensions.
We shall show in Sections II and III that this can indeed be
achieved. Moreover, unlike the Einstein-Hilbert action or the
Samuel-Jacobson-Smolin action, the Euclidean action of this theory is
{\it positive semi-definite for arbitrary four-topologies}.
This positive semi-definite Euclidean action can be constructed by
adding to the Samuel-Jacobson-Smolin action appropriate Yang-Mills
and topological terms of the (anti)self-dual connection$-$terms which
are naturally associated with actions based upon fundamental gauge
connections.

A positive semi-definite Euclidean action may allow for a well-defined path
integral formulation of the quantum theory in the same spirit of more
familiar Euclideanized Yang-Mills and scalar field quantum theories. This
situation is to be contrasted with the proposed Euclidean path integral
formulation of quantum gravity based upon the Einstein-Hilbert
action \cite{haw}. Unlike path integrals in ordinary Euclideanized quantum
field theories, here the Einstein-Hilbert action is not bounded from below;
although it is possible to achieve convergence of the functional integral by
doing a formal conformal rotation, and then performing a suitable contour
integral over complex conformal factors \cite{haw}. These extra manipulations
are however not needed for more familiar theories with {\it positive
semi-definite} Euclidean actions such as Yang-Mills theories.
Moreover, the conformal manipulations achieve formal convergence by
starting from a manifestly divergent kinematic formulation which is based
on a quantity which is not well-defined$-$the
Euclidean gravitational functional integral of the Einstein-Hilbert action
 \cite{schl}.

After a discussion on the equations of motion in Section III, we
consider the Lorentzian action and its analytic continuations in Section
IV. We observe that the proposed action and the Samuel-Jacobson-Smolin action
contain only projections of the curvature
and veirbein combinations which are {\it doubly (anti)self-dual}.
In spinorial terms, this means that analytic continuations of the action
can be phrased {\it solely} in terms of continuations from primed spinors to
primed spinors in complexified spacetimes.

It is also the purpose of this work to give a self-contained discussion
of the effects of discrete transformations C, P and T on the
Samuel-Jacobson-Smolin action and others which employ self- or
anti-self-dual variables as fundamental variables in the description of
four-dimensional gravity. The Einstein-Hilbert action is C, P and T
invariant. Although the Samuel-Jacobson-Smolin action gives rise to the
same equations of motion, in the first order formulation, the fundamental
variables are nevertheless {\it independent} (anti)self-dual connection and
vierbein fields. Since self- or anti-self-dual combinations are not
P-invariant, this raises the possibility that in the first order
formulation, gravity described in terms of these variables is {\it
off-shell} P-non-conserving, despite the fact that the equations of motions
are the same as Einstein's. There are further issues at stake. Naively, CPT is
expected to be good since it is well-known by the CPT theorem that
any Lorentz-invariant {\it hermitian} local action is
CPT invariant. However, there is a subtlety here, due to the
fact that in Lorentzian signature spacetimes, self- or anti-self-dual
connections are actually {\it complex} combinations$-$for
instance, $A^{\pm}_{BC} ={1\over 2}(\star A_{BC} \pm iA_{BC})$.
Thus it can happen that although local Lorentz transformations
remain as symmetries of the theories, actions based on such variables may
not be hermitian, and can contain anti-hermitian or pure imaginary
Lorentz-invariant local pieces. Such terms are CPT {\it odd}.
These, and various other issues connected with the discrete transformations
C, P and T are discussed in the last section.

\section*{II. Positive semi-definite Euclidean Action}

Let us briefly recall some concepts with regard to duality and establish
our notations.
If a two-form carries a pair of anti-symmetric internal
indices AB, with each index taking values from 0 to 3; it is
possible to consider the notion of both internal {\it and} external self-
or anti-self-duality.
The curvature two-form, ${1\over 2}R_{AB\mu\nu}dx^\mu\wedge dx^\nu$,
of the spin connection in four-dimensions is an example.
On two-forms, the internal dual transformation is defined to be
\begin{equation}
\star C_{AB\mu\nu}dx^\mu\wedge dx^\nu \equiv {1\over 2}\epsilon_{AB}\,^{CD}
C_{CD\mu\nu}dx^\mu\wedge dx^\nu
\end{equation}
and the external dual transformation
\begin{equation}
*C_{AB\mu\nu}dx^\mu\wedge dx^\nu \equiv {1\over 2}|e|
\epsilon^{\mu\nu}\,_{\alpha\beta} C_{AB\mu\nu}dx^\alpha \wedge dx^\beta
\end{equation}
The internal (external) indices are raised and lowered by $\eta^{AB}$
($g^{\mu\nu}$)  and $\eta_{AB}$ ($g_{\mu\nu}$) respectively, and $|e|$ is
the determinant of the vierbein, $e^A = e^A\,_\mu dx^{\mu}$.
We shall consider only the case for which the internal indices are Lorentz
indices, and the signatures of the internal and external
metrics are the same i.e. $(\mp, +,+,+)$. We adopt the
convention that upper case Latin indices which run from 0 to 3 denote
Lorentz indices, while lower case indices run from 1 to 3
e.g. A = 0, a ;  a = 1,2,3.

An interesting special case is the two-form $\Sigma_{AB} = e_{A} \wedge
e_{B}$. For it, external and internal dual transformations are the same since
\begin{eqnarray}
*\Sigma_{AB} &=& {1\over 2}|e|
\epsilon^{\mu\nu}\,_{\alpha\beta} e_{A\mu} e_{B\nu} dx^\alpha \wedge dx^\beta
\nonumber\\
&=&{1\over 2}
\epsilon_{AB}\,^{CD} e_{C\alpha} e_{D\beta} dx^\alpha \wedge dx^\beta
=\star \Sigma_{AB}
\end{eqnarray}
Since $\star^2 = *^2 = \pm1$, the eigenvalues are $\pm{1}$ and $\pm{i}$
for Euclidean and Lorenztian signature respectively.
Internal self- and anti-self-dual combinations for Lorentzian signature are
denoted as
\begin{equation}
G^\pm_{AB} = {1\over 2}(\star G_{AB} \pm{i}G_{AB})
\end{equation}
These satisfy
\begin{equation}
\star G^{\pm}= {\pm}iG^{\pm}
\end{equation}
Note that $G_{AB}$ does not have to be a two-form for
{\it internal} self or anti- self-duality to make sense. For instance, one
can consider the anti-self-dual combination of the connection one-form
$A^-_{BC} ={1\over 2}(\star A_{BC} - {i}A_{BC})$. The combinations
${1\over 2}(*C_{AB} \pm{i}C_{AB})$ are external self- and anti-self-dual
two-forms.
In the above self and anti-self-dual combinations, the $i$'s should be
set to unity for the case of Euclidean signature.

We shall start with the proposed positive semi-definite Euclidean action and
consider the Lorentzian case later on.
It is a matter of convention to use either self-dual or anti-self-dual
variables. We choose to use anti-self-dual variables for all our discussions.
Our conventions will then be that anti-self-dual variables are
coupled to left-handed fermions fields.

Let $A_{BC} = -A_{CB}$ be a connection one-form, and $A^-_{BC}$ be the
anti-self-dual combination $A^-_{BC} = {1\over 2}(\star A_{BC}- A_{BC})$.
It can be verified that the curvature of two-form of $A^-_{BC}$,
\begin{equation}
F^-_{AB} = dA^-_{AB} + A^-_{AC}\wedge A^{-C}\,_B
\end{equation}
satisfies
\begin{equation}
F^-_{AB} = {1\over 2}(\star F_{AB} -  F_{AB})
\end{equation}
where $F_{AB} = dA_{AB} + {1\over 2}A_{AC}\wedge A^C\,_B$.

The proposed action is
\begin{eqnarray}
S_E = &-&\int_M \left[{1\over{2g}}(F^-_{AB} - *F^-_{AB})-
{g\over{(16\pi{G})}} \Sigma^-_{AB}\right]\wedge
\left[{1\over{2g}}(F^{-AB} - *F^{-AB})- {g\over{(16\pi{G})}}
\Sigma^{-AB}\right]
\nonumber\\
=&-&\int_M [{1\over{2g^2}}(F^-_{AB}\wedge F^{-AB} -
*F^-_{AB}\wedge F^{-AB}) - {1 \over 8\pi{G}}F^-_{AB}\wedge\Sigma^{-AB}
\nonumber\\
&+&{g^2\over{(16\pi{G})^2}}{\Sigma^-_{AB}}\wedge{\Sigma^{-AB}}]
\end{eqnarray}

The combination
$\left[{1\over{2g}}(F^-_{AB} - *F^-_{AB}) - {g\over{(16\pi{G})}}
\Sigma^-_{AB}\right]$ is anti-self-dual under {\it both}
external and internal duality transformations.\footnote{Note that
$\star \Sigma = *\Sigma$ implies that $*\Sigma^- = {1\over
2}*(\star\Sigma -\Sigma) = -\Sigma^-$.}
If all the variables
and couplings are real for Euclidean signature, then the Euclidean action
is {\it positive semi-definite} for {\it arbitrary topologies} because the
integrand in (9) is, since the action is also
\begin{equation}
S_E = \int_M
\left[{1\over{2g}}(F^-_{AB} - *F^-_{AB})-{g\over{(16\pi{G})}}
\Sigma^-_{AB}\right]\wedge
*\left[{1\over{2g}}(F^{-AB} - *F^{-AB})-{g\over{(16\pi{G})}}
\Sigma^{-AB}\right]
\end{equation}
Recall that for Riemannian spacetimes, the inner product for differential
forms, $({\bf \alpha},{\bf \beta})=\int_M{\bf \alpha}\wedge*{\bf \beta}$,
leads to $(\alpha,\alpha) \geq 0$.

Let us examine the terms in the action $S_E[e^A, A^-_{AB}]$.
The first term within brackets in the second line of (9) is a locally exact
topological term which does not contribute to the equations of motion.
Locally, it can be written in terms of the Chern-Simons \cite{cs} three-form
${\cal C}^-$ of the Ashtekar variables as $d{\cal C}^-[A^-_{AB}]$.
\footnote{The Ashtekar variables can be assumed to be
just $A^-_{0a}$ since $A^-_{bc}$ and $A^-_{0a}$ are not independent,
but are related by
$A^-_{bc} = i\epsilon^{0a}\,_{bc}{A^-_{0a}}$. Again, for Euclidean
signature, the $i$ here should be set to unity.}
It can also be expressed in the form of topological Euler and
signature invariants since
\begin{equation}
\int_M F^-_{AB} \wedge F^{-AB} =
\int_M [{1\over 2}F^{AB}\wedge F_{AB} +
{1 \over 4}\epsilon_{ABCD}F^{AB} \wedge F^{CD}]
\end{equation}
while for compact four-manifolds without boundary, the signature and Euler
invariants are respectively
\begin{equation}
\tau(M) = -{1\over{24\pi^2}}\int_M R^{AB}\wedge R_{AB}
\end{equation}
and
\begin{equation}
\chi(M) = {1 \over{32\pi^2}}\int_M \epsilon_{ABCD}R^{AB}\wedge R^{CD}
\end{equation}
The remaining term is the Yang-Mills action $\int_M Tr( F^-\wedge*F^-)$
for the Ashtekar fields.

Next is
\begin{equation}
{1\over 8\pi{G}}\int_M F^-_{AB}\wedge\Sigma^{-AB} =
{1\over 16\pi{G}}\int_M\left[F_{AB}\wedge e^A\wedge e^B - *F_{AB}\wedge
e^A\wedge e^B\right]
\end{equation}
This is the Samuel-Jacobson-Smolin
action \cite{js, sam} for the Ashtekar variables.
The second term on the RHS of the above is the Einstein-Hilbert-Palatini
action.

The last entry in (9) is just the cosmological term since
\begin{equation}
\Sigma^-_{AB}\wedge\Sigma^{-AB} = -3!e^0\wedge e^1\wedge e^2\wedge e^3=-6(*1)
\end{equation}

By comparing the coupling constants, the (positive) cosmological constant
is related to $g$ and the gravitational constant, $G$, by
\begin{equation}
g^2 = {16\pi\lambda G \over 3}
\end{equation}

Putting everything together, the total Lagrangian in (9) corresponds to
adding a topological term as well as a non-topological
Yang-Mills $Tr(F^-\wedge *F^-)$
Lagrangian four-form to the Samuel-Jacobson-Smolin Lagrangian with
cosmological constant.

A related action without the Yang-Mills term was proposed recently by
Nieto {\it et al} \cite{nieto} in the context of an (anti)self-dual version
of the SO(3,2)
MacDowell-Mansouri action \cite{man} for gravity. Their action can also be
written as
\begin{equation}
S = -\int_M
\left[{1\over{g}}F^-_{AB} - {g\over{(16\pi{G})}}
\Sigma^-_{AB}\right]\wedge
\left[{1\over{g}}F^{-AB} - {g\over{(16\pi{G})}}
\Sigma^{-AB}\right]
\end{equation}
This particular action leads to exactly the same equations of motion as
Einstein's since it differs from the Samuel-Jacobson-Smolin action with
cosmological constant by only a topological invariant. However, since
$F^-$ is not {\it externally} anti-self-dual {\it off-shell}, the Euclidean
action is not positive semi-definite, unlike the proposed action (9). In
(9) only {\it doubly} (anti)self-dual fluctuations of the curvature and
$\Sigma$ contribute to the action.
It is intriguing to observe that this is also true for the
Samuel-Jacobson-Smolin action which can be rewritten as
${1\over 16\pi{G}}\int_M(F^-_{AB} - *F^-_{AB})\wedge\Sigma^{-AB}$
because of the anti-self-dual nature of $\Sigma^-$ with respect to $*$.
However, this action is again not positive definite. In this respect, it
is quite natural to view the proposed action (9) (or (10)) as the natural
positive definite extension which preserves the condition that only doubly
(anti)self-dual fluctuations of the curvature and $\Sigma$ contribute.
As we shall discuss later in Section IV, the fact that the action contains
only doubly anti-self-dual projections means that, in spinorial terms, when
spacetimes are complexified, analytic continuations of the action can be
phrased solely in terms of continuation from primed to primed spinors.

An alternative to the action (10) which also satisfies the
criterion of positive semi-definite action is
\begin{equation}
S = -\int_M
\left[-{1\over{g}}F^-_{AB} + {g\over{(16\pi{G})}}
\Sigma^-_{AB}\right]\wedge
*\left[-{1\over{g}}F^{-AB} + {g\over{(16\pi{G})}}
\Sigma^{-AB}\right]
\end{equation}
Although this latter action contains both (externally) self {\it and}
(anti)self-dual projections of the curvature $F^-$, the equations of motion
obtain from (10) and (18) are the same because they differ only by
a topological term proportional to $\int_M Tr(F^-\wedge F^-)$.

\section*{III. (Anti)self-duality and Einstein Manifolds}

We turn next to the equations of motion of the proposed
actions. We shall show that {\it all} Einstein manifolds are solutions to
the
equations of motion of the new actions. Furthermore, $F^-$ is
(anti)self-dual
(with respect to $*$) iff the solution is an Einstein manifold.
\footnote{Usually, self- or anti-self-duality of Yang-Mills gauge
fields refers to self- or anti-self-duality of the curvature with respect to
$*$. This is what is being discussed here.
The difference here is that we are also using
variables which are also internally (anti)self-dual.}
Thus, {\it the on-shell (anti)self-dual sector of the theory corresponds
precisely to all Einstein manifolds in four dimensions.}

Varying the first order action $S_E[e^A, A^-_{AB}]$ with respect to $A^-$
yields the equation of motion
\begin{equation}
-{1\over g^2}D_{A^-}*F^- + {1\over 16\pi{G}}D_{A^-}\Sigma^- = 0
\end{equation}
The additional metric dependent Yang-Mills term contributes to the
energy-momentum tensor.
So varying with respect to $e^A\,_\mu$ produces an extra contribution.
These equations of motion come from
\begin{eqnarray}
\delta S_E|_{A^-}&=&\int_M -{1\over 8\pi{G}}
(F^-_{AB}\wedge e^B\wedge
dx^\mu + {\lambda \over 3!}\epsilon_{ABCD}e^B\wedge e^C\wedge e^D
\wedge dx^\mu)\delta e^A_\mu
\nonumber\\
&&+ {2 \over g^2}\int_M T_{YM}^{\mu\nu} e_{A\nu} \delta e^A_\mu (*1)
\nonumber\\
&=& 0
\end{eqnarray}
Here $T_{YM}^{\mu\nu}$ is the energy-momentum tensor from the Yang-Mills
Lagrangian. It takes the form
\begin{eqnarray}
T_{YM}^{\mu\nu} &=&{1\over 4}g^{\mu\nu}F^-_{AB\alpha\beta}F^{-AB\alpha\beta}-
{F^-_{AB}}^\mu\,_\alpha {F^{-AB\nu\alpha}}
\nonumber\\
&=& -{1\over 2}({F^-_{AB}}^\mu\,_\alpha{F^{-AB\nu\alpha}}
\mp *{F^-_{AB}}^\mu\,_\alpha *F^{-AB\nu\alpha})
\end{eqnarray}
In arriving at the last equality we have used the identity
\begin{equation}
{1\over 2}g^{\mu\nu}F^-_{AB\alpha\beta}F^{-AB\alpha\beta}-
{F^-_{AB}}^\mu\,_\alpha{F^{-AB\nu\alpha}}
= \pm *{F^-_{AB}}^\mu\,_\alpha *{F^{-AB\nu\alpha}}
\end{equation}
where the upper(lower) sign is for Euclidean(Lorentzian) signature.
{}From (21) it is clear that the energy-momentum tensor vanishes if the
curvature is self- or anti-self-dual.\footnote{For Lorentzian
signature, the eigenvalues of $*$ are $\pm{i}$. So the result that the
energy-momentum from the Yang-Mills action vanishes for self- or
anti-self-dual curvatures ${\cal F}$ which satisfy $*{\cal F} =
\pm{i}{\cal F}$, also holds for Lorentzian signature.}
This will be used to show that {\it all} Einstein manifolds are
solutions to the equations of motion of the proposed actions.

It is known that the Einstein field equations in four-dimensions can be
obtained from the Samuel-Jacobson-Smolin action with cosmological constant.
In terms of $A^-$ and $e$, they read
\begin{equation}
D_{A^-}\Sigma^- = 0
\end{equation}
and
\begin{equation}
F^-_{AB}\wedge e^B\wedge + {\lambda \over 6}\epsilon_{ABCD}e^B\wedge e^C
\wedge e^D = 0
\end{equation}
Given an invertible veirbein, the unique solution to (23) is
that
$A^-$ is the (anti)self-dual part of the torsionless spin connection i.e.
\begin{equation}
A^-_{AB} = {1\over 2}(\star\omega_{AB}(e) - \omega_{AB}(e))
\end{equation}
which makes the curvature $F^-_{AB} = {1\over 2}(\star R_{AB} - R_{AB})=
R^-_{AB}(e)$. Solutions that satisfy this and (24) are Einstein manifolds.
For veirbeins of Einstein manifolds, the Ashtekar connections
$A^-_{Einstein}$ as in
(25) are (externally) (anti)self-dual as well i.e. $F^- = -*F^-$.
\footnote{Recall that the Riemann-Christofel curvature
tensor for Einstein manifolds in four dimensions, $R_{ABCD}$,
obey the condition that the left and right dual are
equal. So for $R_{AB} = {1 \over 2}R_{ABCD}e^C\wedge e^D$ , $\star R_{AB}
= *R_{AB}$. Consequently, $*R^-_{AB}= -R^-_{AB}$.}
As noted previously, the extra energy-momentum tensor contribution from the
Yang-Mills term in (20) vanishes for self or anti-self-dual curvatures.
So Ashtekar connections and veirbeins for Einstein
manifolds also obey the set of equations (19) and (20). Conversely,
if the connection $A^-_{AB}$ is such that $F^-_{AB} = -*F^-_{AB}$,
then the set (19) and (20) reduces to the set (23) and (24) due to the
Bianchi identity $D_{A^-}F^- = 0$, and the vanishing of the
Yang-Mills energy-momentum tensor. Therefore we can conclude that, on-shell,
the (anti)self-dual sector of the new actions (10) and (18) corresponds
{\it precisely} to all Einstein manifolds in four dimensions.

Expression (10) tells us that the positive semi-definite Euclidean action
$S_E$ is minimized by configurations which obey
\begin{equation}
{1\over 2}(F^- - *F^-) = {g^2 \over {(16\pi{G})}}\Sigma^- =
{\lambda\over3}\Sigma^-
\end{equation}
For Einstein manifolds, the Weyl two-form is given by
\begin{equation}
W_{AB} = R_{AB} -{\lambda\over 3}e_A\wedge e_B
\end{equation}
and $*R_{AB} = \star R_{AB}$. The latter leads to
$F^-_{AB}(A^-_{Einstein})$$= -*F^-_{AB}$$(A^-_{Einstein})$.
The anti-self-dual part of the Weyl two-form is therefore
\begin{eqnarray}
W^-_{AB} &=& {1\over 2}(\star W_{AB} - W_{AB})
\nonumber\\
&=& {1\over 2}(F^-_{AB} - *F^-_{AB}) - {g^2 \over 16\pi{G}}\Sigma^-_{AB}
\end{eqnarray}
Thus, the lower bound of zero action is attained for
conformally self-dual $(W^-_{AB} =0)$ Einstein manifolds.
Configurations which obey (26) may correspond to
the ground state of the theory.

\section*{IV. The Lorentzian action and analytic continuation}

What is the Lorentzian signature action, $S_L$, which corresponds to the
positive semi-definite action $S_E$ in (10)? We would like the
continuation from Lorentzian to Euclidean signature to have certain
properties. In particular, the actions should have the property that
$exp(iS_L) = exp(-S_E)$.
The continuation should also preserve the (anti)self-dual nature of the
fields $A^-$, $F^-$ and $\Sigma^-$ with respect to $\star$
\footnote{Our convention for Lorentzian signature is
$\epsilon_{0123}=-\epsilon^{0123}=1$.},
as well as the (anti)self-duality of the combination ${1\over 2}(*F^- -
(i)F^-)$  and $\Sigma^-$ with respect to $*$.
We shall first show explicitly, that it is possible to achieve these and
continue from Lorentzian to Euclidean signature and vice versa by a
Wick rotation before giving a more general analytic continuation
prescription.

The Lorentzian action is
\begin{eqnarray}
S_L &=&-\int_M \left[{1\over{2g}}(F^-_{AB} + i*F^-_{AB})-
{g\over{(16\pi{G})}} \Sigma^-_{AB}\right]\wedge
*\left[{1\over{2g}}(F^{-AB} + i*F^{-AB})- {g\over{(16\pi{G})}}
\Sigma^{-AB}\right]
\nonumber\\
&=& i\int_M [{1\over{2g^2}}(F^-_{AB}\wedge F^{-AB} +
i*F^-_{AB}\wedge F^{-AB}) - {1 \over 8\pi{G}}F^-_{AB}\wedge\Sigma^{-AB}
\nonumber\\
&&+{g^2 \over {(16\pi{G})}^2}\Sigma^-_{AB}\wedge\Sigma^{-AB}]
\end{eqnarray}

In this section and henceforth, unless stated otherwise, all variables
are Lorentzian. To be clear, these may be denoted with $L$
subscripts. Euclidean variables will be denoted by $E$ subscripts.

A Wick rotation with $(e_L)_0 = -i(e_E)_0$ and $(e_L)_a = (e_E)_a$ will
result in the metric having Euclidean signature $(+, +, +, +)$.
The corresponding change induced in $\Sigma^-$ is
\begin{equation}
(\Sigma^-_L)_{0a} \mapsto (\Sigma^-_E)_{0a} = {1\over 2}({1\over
2}\epsilon_{0a}\,^{bc} (\Sigma_E)_{bc} - (\Sigma_E)_{0a})
\end{equation}
Thus
\begin{equation}
\Sigma^{-0a}_L= -(\Sigma^-_L)_{0a} \mapsto -(\Sigma^-_E)_{0a} =
-\Sigma^{-0a}_E
\end{equation}
We also have $det(e^A_\mu)_L \mapsto idet(e^A_\mu)_E$.
With
\begin{eqnarray}
(A_L)_{0a} \mapsto -i(A_E)_{0a}
\nonumber\\
(A_L)_{bc} \mapsto (A_E)_{bc}
\end{eqnarray}
we obtain
\begin{equation}
(A^-_L)_{0a} \mapsto (A^-_E)_{0a} = {1\over 2}({1\over
2}\epsilon_{0a}\,^{bc} (A_E)_{bc} -(A_E)_{0a})
\end{equation}
For the curvature,
\begin{equation}
(F^-_L)_{0a} \mapsto (F^-_E)_{0a} = {1\over 2}({1\over 2}\epsilon_{0a}\,^{bc}
(F_E)_{bc} -(F_E)_{0a})
\end{equation}
To render the continuation explicit, we note that
$\Sigma^-_{bc} = (i)\epsilon_{bc}\,^{0a}\Sigma^-_{0a}$ and
$F^-_{bc} = (i)\epsilon_{bc}\,^{0a}F^-_{0a}$. The action is therefore
\begin{eqnarray}
S_L&=& i\int_M [{1\over{2g^2}}(F^-_{AB}\wedge F^{-AB} +
i*F^-_{AB}\wedge F^{-AB}) - {1 \over 8\pi{G}}F^-_{AB}\wedge\Sigma^{-AB}
\nonumber\\
&&+{g^2 \over {(16\pi{G})}^2}\Sigma^-_{AB}\wedge\Sigma^{-AB}]
\nonumber\\
&=&i\int_M [{4\over{2g^2}}(F^-_{0a}\wedge F^{-0a} +
i*F^-_{0a}\wedge F^{-0a}) - {4 \over 8\pi{G}}F^-_{0a}\wedge\Sigma^{-0a}]
\nonumber\\
&&+{4g^2 \over {(16\pi{G})}^2}\Sigma^-_{0a}\wedge\Sigma^{-0a}]
\end{eqnarray}
Thus
\begin{eqnarray}
iS_L = &&\int_M [{4\over{2g^2}}(F^-_{0a}\wedge(-F^{-0a}) +
i*F^-_{0a}\wedge(-F^{-0a})) - {4 \over 8\pi{G}}F^-_{0a}\wedge(-\Sigma^{-0a})
\nonumber\\
&+&{4g^2 \over {(16\pi{G})}^2}\Sigma^-_{0a}\wedge(-\Sigma^{-0a})]
\end{eqnarray}
is continued to
\begin{eqnarray}
&&\int_M [{4\over{2g^2}}((F^-_E)_{0a}\wedge F^{-0a}_E
- *_E(F^-_E)_{0a}\wedge
F^{-0a}_E) - {4 \over 8\pi{G}}(F^-_E)_{0a}\wedge\Sigma^{-0a}_E
\nonumber\\
&&+{4g^2 \over {(16\pi{G})}^2}(\Sigma^-_E)_{0a}\wedge\Sigma^{-0a}_E]
\nonumber\\
&=&\int_M [{1\over{2g^2}}((F^-_E)_{AB}\wedge F^{-AB}_E
- *_E(F^-_E)_{AB}\wedge F^{-AB}_E) - {1 \over 8\pi{G}}
(F^-_E)_{AB}\wedge\Sigma^{-AB}_E
\nonumber\\
&&+{g^2 \over {(16\pi{G})}^2}(\Sigma^-_E)_{AB}\wedge\Sigma^{-AB}_E]
\nonumber\\
&=&-S_E
\end{eqnarray}
where $S_E$ is precisely as in expression (9). So we indeed have a
continuation of $exp(iS_L)$ to $exp(-S_E)$ with positive semi-definite
Euclidean action $S_E$.

Although the Euclidean action (9) is positive semi-definite for arbitrary
topologies, it remains to be seen whether
this can provide all the necessary convergence properties for a well-defined
Euclidean path integral approach to quantum gravity. The actions (9),
(17) and (18) also contain a dimensionless coupling constant,
$g = \sqrt {{16\pi\lambda{G}}\over 3}$, but the perturbative
renormalizability (or non-renormalizability) of these theories has not yet
been studied.

The action can be thought of as $S_L[e^A, A^-_{0a}]$.
In the explicit example of continuation from Lorentzian to Euclidean
signature, we Wick rotated only the imaginary part of $A^-$.
In continuing from Euclidean to Lorentzain signature however, we have to be
careful since $A^-$ is real in Euclidean signature spacetimes.
How then can we distinguish which part of $A^-$ to Wick rotate which to
leave invariant\footnote{A common concept for continuing between Lorentzian
and Euclidean signature and vice versa is that we should Wick rotate the parity
odd part of $A^-$ and leave the parity even part of $A^-$ unchanged.
We shall have more to say on the properties of $A^-$ under parity in the
next section.} if we insists on using solely (anti)self-dual
variables? In general, what we are actually seeking is a continuation
which preserves the (anti)self-dual nature of the fields.
Precisely, in spinorial terms, the relevant analytic continuations that
we seek are continuations {\it from primed spinors to primed spinors}.

In complexifying spacetimes, quantities which are complex on Lorentzian
sections must be treated as independent of their complex
conjugates\cite{haw}. This is the case for fermions and scalar
fields in usual Euclideanized quantum field theories.
Thus $A^-$ and $A^+$, which are complex conjugates of
each other in Lorentzian signature spacetimes, have to be analytically
continued to different independent fields in complex spacetimes.
When continued to the Euclidean section these are again
independent.

In spinorial terms, ${A^-}_{{\cal A}'}\,^{{\cal B}'} = A^-_{0a}{1\over\sqrt
2}({\tau^a})_{{\cal A}'}\,^{{\cal B}'}$ (the scripted spinorial indices take
values 0 and 1; and $\tau^a$ are Pauli matrices), is a primed
(left-handed or dotted) spinor. Note that here $A^-_{0a}$ is a one-form, with
components $A^-_{0a\mu}$ which can also be expressed in primed {\it and
unprimed} spinorial indices by contracting with curved-space spinors
$\sigma^\mu_{{\cal A}{\cal A}'}.$
\footnote{$\sigma^\mu_{{\cal A}{\cal A}'}$ are ``soldering" spinors which
satisfy $g_{\mu\nu}\sigma^\mu_{{\cal A}{\cal A}'}\sigma^\nu_{{\cal
B}{\cal B}'} = \epsilon_{{\cal A}'{\cal B}'}\epsilon_{{\cal A}{\cal B}}$.}
This contraction can be done for each external or spacetime index of any
tensor. In particular, it can be done for the components of the
{\it externally} (anti)self-dual objects, $\Sigma^-$ and
${1\over2}(*F^- - (i)F^-)$.\footnote{Again, the $(i)$ should be set to
unity for Euclidean signature.} However, we may note that combinations
such as $\Sigma^-_{\mu\nu}$ and ${1\over2}(*F^-_{\mu\nu}-(i)F^-_{\mu\nu})$
are {\it also internally (anti)self-dual}; and therefore in spinorial
terms, contain {\it only primed projections} due to their {\it doubly}
(anti)self-dual nature. The action consists of only these projections,
and can thus be written {\it exclusively in terms of primed spinors}.
\footnote{As we have mentioned before, this is also a
property of the Samuel-Jacobson-Smolin action.}
As a result, the Lorentzian-Euclidean continuation can be phrased in the
more general and rigorous context of analytic continuations of primed
spinors to primed spinors in complex spacetimes.

\section*{V. Discrete transformations C, P and T; and (anti)self-dual
variables}

In this section we discuss the charge conjugation (C), parity (P) and
time reversal (T) transformations and their combinations; and examine the
effects of these discrete transformations on the theory.

To give a coordinate independent description, it is convenient to use
differential forms. P and T are improper Lorentz transformations of
determinant $-1$ and act on the Lorentz indices. In particular
under P
\begin{equation}
e^0 \mapsto e^0 ;\quad e^a \mapsto -e^a
\end{equation}
while under T
\begin{equation}
e^0 \mapsto -e^0 ;\quad e^a \mapsto e^a
\end{equation}
Both P and T are {\it orientation-reversing}($M \rightarrow
{\overline M}$) operations but unlike P,
which is to be implemented unitarily, T is to be implemented by anti-unitary
transformations. So under T, c-numbers are complex conjugated.
C acts trivially on
the veirbein. On the spin connections, $\omega$, the induced
transformations can be deduced from the torsionless condition,
\begin{equation}
de^A + \omega^A\,_B \wedge e^B = 0
\end{equation}
and they are such that under both P and T,
\begin{equation}
\omega_{0a} \mapsto -\omega_{0a} ;\quad \omega_{bc} \mapsto \omega_{bc}
\end{equation}
The induced transformations on the curvature, $R_{AB}= d\omega_{AB} +
\omega_A\,^C \wedge \omega_{CB} $,
are that
\begin{equation}
R_{0a} \mapsto - R_{0a} ;\quad R_{bc} \mapsto R_{bc}
\end{equation}
For Lorentzian signature, the anti-self-dual and self-dual combinations
$\omega^{\mp}_{0a} \equiv {1\over 2}({\pm}i\omega_{0a}+{1\over
2}\epsilon_{0a}\,^{bc} \omega_{bc})$
behave as
\begin{equation}
\omega^-_{0a} \leftrightarrow \omega^+_{0a}
\end{equation}
under P, while under T (which is anti-unitary)
\begin{equation}
\omega^{\mp}_{0a} \leftrightarrow \omega^{\mp}_{0a}
\end{equation}
$\omega^\pm$ are trivially invariant under C.

All the terms in the action are  invariant under local $SO(3,C)$
gauge transformations
and diffeomorphisms. $F^-$ and $\Sigma^-$ transform covariantly while
$A^-$ is an (anti)self-dual (left-handed) connection, and a singlet under
right-handed Lorentz transformations. With the Samuel-Jacobson-Smolin
action, the Ashtekar connection is indeed the (anti)self-dual combination of
the spin connection in the second order formulation.
So it is reasonable to assume that the behaviour of
the Ashtekar connections under the discrete transformations is the
same as the behaviour of the (anti)self-dual part of the spin connection.
This is in agreement with the fact that the connection carries Lorentz
indices according to
$A^-_{AB} = {1\over2}({1\over2}\epsilon_{AB}\,^{CD}A_{CD}-iA_{AB})$.

In order to verify that these properties of $A^\pm$ under discrete
transformations are indeed correct, and are compatible with the usual
notions of P and T; we can couple
fermions to the theory and consider the invariance of the Dirac Lagrangian
or the Dirac equation.\footnote{Since the Ashtekar-Sen connections are
either self- or anti-self-dual, fermions of only one
chirality can be coupled to either of these connections. However, it is
possible to write the bispinor Dirac equation and Dirac Lagrangian in terms
of a pair of left-handed Weyl fermions by substituting
$\phi_R = -i\tau^2(\chi_L)^*$ \cite{art, cc}. For a discussion of
anomaly-free fermion couplings to Ashtekar-Sen connection, the effects of
discrete transformations, and whether such couplings can produce
the phenomenology of the Standard Model and Beyond,
see Ref. \cite{cc}.}

The massless Dirac equation is
\begin{equation}
{\gamma^A}{E_A}\lfloor D\Psi = 0
\end{equation}
Here $E_A$ are the inverse veirbein vector fields $E^\mu_A \partial_\mu$,
and the contractions are such that $E_A\lfloor e^B = \delta_A\,^B$, $
E_A{\lfloor}D = E^{\mu}_A{D_\mu}$. $\Psi$
is a four-component Dirac bispinor; and in the chiral representation with
\begin{equation}
\gamma^5 =\left(\matrix{I_2 &0 \cr 0 &-I_2\cr}\right)
\end{equation}
the covariant derivative is
\begin{equation}
D\Psi = \left[dI_4 - i\left(\matrix{A^+_a \tau^a/2 &0\cr 0
&A^-_a\tau^a/2\cr}\right)\right] \left(\matrix{\phi_R\cr \phi_L\cr}\right)
\end{equation}
where $\phi_{R,L}$ are two-component right and left-handed Weyl spinors.
$A^{\mp}_a \equiv {\pm}iA_{0a} -{1\over 2}\epsilon_{0a}\,^{bc}
A_{bc} = -2A^\mp_{0a}$. In
terms of Pauli matrices, $\tau^a = -{\overline\tau}^a$, and
$\tau^0 = {\overline\tau}^0 = -I_2$, the Dirac matrices in the chiral
representation are
\begin{equation}
\gamma^A =\left(\matrix{0 &i\tau^A\cr i{\overline\tau}^A &0\cr}\right)
\end{equation}
Note that
\begin{equation}
{1\over 4}A^+_{BC}\tau^B{\overline \tau}^C = (iA_{0a}+ {1\over 2}
\epsilon_{0a}\,^{bc}A_{bc}){\tau^a \over 2}=
-A^+_a{\tau^a\over 2}
\end{equation}
and
\begin{equation}
{1\over 4}A^-_{BC}{\overline \tau}^B\tau^C =
(-iA_{0a}+ {1\over 2}\epsilon_{0a}\,^{bc}A_{bc}){\tau^a \over 2}
=-A^-_a{\tau^a \over 2}
\end{equation}
So (47) is analogous to the usual fermion coupling to spin
connections for which the covariant derivative is
\begin{eqnarray}
D_\omega\Psi &=&(dI_4 + {1\over 8}\omega_{BC}[\gamma^B, \gamma^C])\Psi
\nonumber\\
&=&\left[ dI_4 +
i\left(\matrix{
{{1\over 4}\omega^+_{BC}\tau^B{\overline \tau}^C}
&0\cr 0
&{{1\over 4}\omega^-_{BC}{\overline \tau}^B\tau^C}\cr}\right)
\right] \left(\matrix{\phi_R\cr \phi_L\cr}\right)
\end{eqnarray}

Under P, we have $\Psi \mapsto i\gamma^0\Psi$, so that $\phi_L
\leftrightarrow \phi_R$.\footnote{We shall leave out all the intrinsic
phases since these complications will not come into play in our
discussion of the transformation properties of $A^-$.} Under T
(which is anti-unitary),
$\Psi \mapsto
-i\tau^2\Psi$; while under C, $\Psi \mapsto {\cal C}{\overline\Psi}^T$
where ${\cal C} = -i\gamma^0\gamma^2$.
It is then straightforward to check that for the Dirac equation to
hold under C, P and T; together with (38) and (39), $A^\pm$ have to
transform according to
\begin{equation}
A^-_a \leftrightarrow A^+_a
\end{equation}
under P, and
\begin{equation}
A^{\mp}_a \leftrightarrow A^{\mp}_a
\end{equation}
under T (which is anti-unitary), and be invariant under C.

In order to discuss the effects of the discrete transformations in a clear
and concise manner, and also to compare with conventional actions, it is
better to refer to the transformation of the variables
$A_{BC} = -A_{CB}$ and its curvature components.
Recall that
\begin{equation}
A_{0a}= i(A^-_{0a} - A^+_{0a}) ;\quad A_{bc} = {\epsilon_0}^a\,_{bc} (A^-_a
+ A^+_a)
\end{equation}
The curvature $F_{BC} = dA_{BC} + A_B\,^D \wedge A_{DC}$ then has components
\begin{equation}
F_{0a}= i(F^-_{0a} - F^+_{0a}) ;\quad F_{bc}= {\epsilon_0}^a\,_{bc}
(F^-_{0a} + F^+_{0a})
\end{equation}
Thus
\begin{eqnarray}
A_{0a} \mapsto -A_{0a} ;\quad A_{bc} \mapsto A_{bc}
\nonumber\\
F_{0a} \mapsto -F_{0a} ;\quad F_{bc} \mapsto F_{bc}
\end{eqnarray}
under P and T and are trivially invariant under C.
The torsion is defined to be
\begin{equation}
T^A = de^A + A^A\,_B \wedge e^B
\end{equation}

It can then be shown that the Samuel-Jacobson-Smolin action
with cosmological constant is
\begin{eqnarray}
S_{SJS} = &-&{1\over{(16\pi{G})}}\int_M [e^A\wedge e^B\wedge *F_{AB}
-2\lambda (*1)]
\nonumber\\
&-& {i\over {(16\pi{G})}}\int_M [d(e^A\wedge T_A) - T^A\wedge{T_A}]
\nonumber\\
\end{eqnarray}
and the total action (9) is
\begin{eqnarray}
S_L = &&S_{SJS} + {1\over 4g^2}\int_M[-iF_{AB}\wedge{F^{AB}}+
{1\over2}\epsilon_{ABCD}F^{AB}\wedge F^{CD}]
\nonumber\\
&+&{1\over 4g^2}\int_M[*F_{AB}\wedge F^{AB}+ {i\over
2}\epsilon_{ABCD}*F^{AB}\wedge{F^{CD}}]
\end{eqnarray}

We take the opportunity here to clarify and emphasize some salient features
which are not usually mentioned in the literature with regard to the
Samuel-Jacobson-Smolin action.

The RHS of the first line of (58) is the
Einstein-Hilbert-Palatini action with cosmological constant term.
It can be deduced from the identity (58) that
the Samuel-Jacobson-Smolin action reproduces the same equations of motion as
Einstein's theory.
The equation of motion that is obtained by varying with respect to $A^-$ is
$D_{A^-}\Sigma^- =0$. This has the unique solution that $A^-$ is the
anti-self-dual part of the spin connection, which makes $A^+$ (which is the
complex conjugate of $A^-$) the self-dual part of the spin connection and
$F^{\pm}_{AB} = R^{\pm}_{AB}(\omega(e))$.
The torsion then vanishes {\it on shell}. Therefore the term quadratic in
the torsion, which is {\it not a total divergence}, cannot give rise
to the any extra equations of motion due to its quadratic dependence on
the torsion. So, modulo the
equation of motion $D_{A^-}\Sigma^- = 0$, varying with respect to the
vierbein then reproduces Einstein's equations from the RHS of the first
line of (58).
Remarkably, the first order action $S_{SJS}$ gives the same
equations of motion as Einstein's theory although, without further
conditions on the independent variables, it
cannot even be regarded as being (complex) canonically related to the
Einstein-Hilbert-Palatini action due to the presence of a torsion-squared
term which is not a total divergence.

In the quantum theory, off-shell fluctuations can be expected to contribute.
So it is not clear that the quantum theory from $S_{SJS}$ is the same as
that from Einstein's theory. This is not a bad thing by itself given the
difficulties with quantizing Einstein's theory. However, if one wishes
to be faithful to the latter and still use $S_{SJS}[A^-, e]$ and
(anti)self-dual variables,
one possibility is to strictly impose a condition equivalent to the
vanishing of the torsion {\it off-shell}.
\footnote{When fermion
couplings are included, the torsion-free condition has to be supplemented by
fermionic contributions, but an analogous condition can be imposed \cite{cc}.
However, there can still be subtle discrete symmetry violations due to the
presence of instantons and the Adler-Bell-Jackiw anomaly \cite{cc}.}
This is the analog in the path integral approach of the
reality conditions that have to be imposed on the conjugate variables in the
canonical quantization program \cite{aa}. Otherwise, the
torsion terms lead to an action
which contains {\it imaginary} terms in Lorentzian signature spacetimes,
and can cause {\it off-shell} CPT violation
since purely imaginary (real) local Lorentz-invariant action terms are CPT
odd (even).\footnote{The behaviour of various terms in the action under
C, P and T can be checked using the explicit transformation
properties of the basic variables $A^\pm$ and $e$ discussed previously.}
However, both theories are equivalent on passing to the second order
formulation since in this latter case, $A^-$ is eliminated in terms of the
veirbein,
$F_{AB}$ is replaced by $R_{AB}$, and the non-hermitian torsion terms are
identically set to zero.

It must be said that even if imaginary Lorentz-invariant terms are not
killed by imposing additional conditions in the first order formulation, it
remains to be seen whether the Samuel-Jacobson-Smolin action (or its
extensions which are Euclidean positive semi-definite) can actually give
rise to a successful quantum theory of gravity. Given the difficulties with
a quantum theory based upon the Einstein-Hilbert action, these actions which
contain all classical solutions of the Einstein-Hilbert action may be
worth exploring as alternatives. In this respect, universal P, T
and CPT violations or conservation checks and
experimental signatures \cite{smolin} of these, can be useful
in the evaluation of the various options.

We shall next turn our attention to the other terms in the action (59).
The topological instanton term\footnote{It also occurs in the action (17)
of Nieto {\it et al}, and the $i$ is present in the Lorentzian
case \cite{nieto}.} is
\begin{equation}
{i\over
2g^2}\int_M F^-_{AB}\wedge F^{-AB}
={1\over 4g^2}\int_M[-iF_{AB}\wedge{F^{AB}} +
{1\over2}\epsilon_{ABCD}F^{AB}\wedge F^{CD}]
\end{equation}
The first term is the analog of the signature invariant. However, due to the
presence of $i$; it is P, CP and CPT odd but T even(recall that T is
anti-unitary). To check that this is indeed to be expected, we note that
in the second order formulation, the Samuel-Jacobson-Smolin action with an
additional topological term (60) reduces to
\begin{eqnarray}
S_{second\,order}[e^A]= &-&{1\over{(16\pi{G})}}\int_M [e^A\wedge
e^B\wedge *R_{AB} -\lambda (*1)]
\nonumber\\
&+&{1\over 4g^2}\int_M [-iR^{AB}\wedge R_{AB} +
{1 \over 2}\epsilon_{ABCD}R^{AB} \wedge R^{CD}]
\end{eqnarray}
Thus, the second order action differs from the Einstein-Hilbert-Palatini
form by precisely two topological invariants which correspond to $\tau(M)$
and $\chi(M)$ (see equations (12) and (13) in Section I). The metric is real
but observe that in the second order action,
the term associated with $R^{AB}\wedge R_{AB}$ is pure imaginary.
It can be checked from our earlier analysis in this section that $\int_M
(R^{AB}\wedge R_{AB})$ is P, T odd hence PT even, while
$\int_M (\epsilon_{ABCD}R^{AB} \wedge R^{CD})$ is P, T even. Since T is
anti-unitary, this means that, due to the $i$, the action (61) is not P, CP
and CPT conserving iff $\int_M (R^{AB}\wedge R_{AB})$ does not vanish,
although the equations of motion are identical to Einstein's. This
means that in this case gravitational instantons with
non-vanishing signature invariant, $\tau$, should give rise to P, CP as
well as CPT violations. (T is however not violated in this manner due to
its anti-unitary nature). These violations
cannot be cured by making $g^2$ imaginary,\footnote{g is needed to be
real on the Euclidean section for the Euclidean action to be positive
semi-definite.} since the term associated
with $\chi$ would then be T and CPT odd, while the term associated with
$\tau$ would be P, T odd. The root of the strange
behaviour and violations under the discrete transformations lies in the
anti-self-dual nature of $A^-$.
The complex {\it connection} $A^-$, which carries Lorentz indices
according to $A^-_{AB} = {1\over 2}({1\over2}\epsilon_{AB}\,^{CD}A_{CD}
-{i}A_{AB})$, is neither even nor odd under P, but has even and odd parts
due to its (anti)self-dual nature.
In terms of spinors, $A^-$ is a left-handed (primed or dotted) connection.
We have $A^- \leftrightarrow A^+$ under P.\footnote{In usual Yang-Mills
theories with {\it real} connections, it is the
self- and anti-self-dual {\it curvature} combinations
$\pm i{\vec E} + {\vec B}$ which are neither even nor odd under P,
but has even and odd parts.}
As a consequence, the combination $\int_M Tr(F^- \wedge F^-)$ is neither
even nor odd under P, and is not real even after reality conditions
are taken into account$-$for instance by going to the second order formulation.

It should be emphasized that the usual $\theta$-angle action in non-abelian
gauge theories with hermitian gauge curvature ${\cal F}$ is of the
form $\theta \int_M Tr({\cal F}\wedge {\cal F})$ for Lorentzian
signature,\footnote{$\theta$ is real and note the absence of $i$.
Furthermore, $\int_M
Tr({\cal F}\wedge{\cal F})$ is odd under both P and T.} and is continued to
the Euclidean action $-i\theta\int_M Tr({\cal F}\wedge{\cal F})$. Similarly,
the usual $\theta$-angle term for gravity is proportional to
$-i\theta\tau$ (for Euclidean signature) and
is $-24{\pi^2}\theta\int_M (R^{AB}\wedge R_{AB})$ for Lorentzian
signature spacetimes.
Note that this latter form has the normal instanton $\theta$-term
behaviour of being P, T, CP odd, and  PT, CPT even; in contradistinction
with the previous form in (60).
To be clear about the $i$'s , under a chiral rotation
$\Psi \mapsto exp(i\alpha\gamma^5)\Psi$, instantons contribute to the
change of $exp(-i\alpha\tau/4)$(note the $i$ associated with $\tau$ for
Euclidean signature) in the Euclidean bispinor fermion measure,
$D{\Psi}D{\overline\Psi}$. This results in the usual
Adler-Bell-Jackiw \cite{abjh} anomaly equation
\begin{equation}
\nabla_\mu j^{\mu5} = -{i\over 192\pi^2}R_{\mu\nu\alpha\beta}
*R^{\mu\nu\alpha\beta}
\end{equation}
for Euclidean signature spacetimes.\footnote{See also Ref. \cite{cc} for
anomaly computations in the context of couplings to Ashtekar-Sen
connections.}

It has been proposed \cite{abal} that there will be an analogous
Yang-Mills instanton $\theta$-angle term in the first order formulation
because of large
$SO(3,C)$ gauge transformations of $A^-$.\footnote{More precisely, in
the canonical formulation, large gauge transformations can come only from
the local pure rotations of the Lorentz group $SO(3,C)$.} If this were
equivalent to
an addition of ${\theta\over 16\pi^2}\int_M F^-_{AB}\wedge F^{-AB}$
to the Lorentzian action as has been suggested \cite{abal}, the additional
$\theta$-term is therefore
\begin{equation}
{\theta\over 16\pi^2}\int_M F^-_{AB}\wedge F^{-AB}=
-{\theta\over 32\pi^2}\int_M [R^{AB}\wedge R_{AB} +
{i\over 2}\epsilon_{ABCD}R^{AB} \wedge R^{CD}]
\end{equation}
if the condition that the Ashtekar connection is the (anti)self-dual part
of the spin connection is enforced. The first signature invariant
$\tau(M)$
contribution has the normal behaviour of a Yang-Mills $\theta$-term of being
P, T odd and CPT even. However, due to the presence of the $i$, this time the
second term (which is an Euler number $\chi(M)$\footnote{Note that the
Euclidean Schwarzschild solution has $\chi = 2, \tau = 0$.}
contribution in Euclidean signature) is P even, T odd, and CPT odd.

The remaining terms in (59) are
\begin{equation}
{1\over 4g^2}\int_M[*F_{AB}\wedge F^{AB}+ {i\over
2}\epsilon_{ABCD}*F^{AB}\wedge {F^{CD}}]
\end{equation}
The first term is the usual Yang-Mills action and it is C, P and T even.
The second is P odd, and T even due to the $i$; and therefore CPT odd.
%
%

It may be that the inner product in the canonical version of the
theory, or the path integral measure in path integral approach,
for the yet unavailable quantum theory can restore CPT invariance and other
discrete symmetries despite the apparent non-conservation of these in the
actions. If so, the observations made here may serve to narrow down the
search for the ``correct" measure or inner product. On the other hand, it
may turn out that these violations and their ramifications are genuine features
which emerge from the use of self- or anti-self-dual fundamental variables
and cannot be compensated for. In either case, it is hoped that this work
may serve to draw attention to the peculiar behaviour of theories with
self- or anti-self-dual variables under discrete transformations, the
possible non-hermiticity of the actions, and some of the physical
implications of utilizing these variables in the
description of classical and quantum gravity in four-dimensions.

\section*{Acknowledgments}

I thank Lay Nam Chang for many useful discussions on self-dual
variables and discrete transformations. I am grateful to Lee Smolin for many
valuable comments and encouragement. I have also benefitted from
discussions with Fernando
Barbero and Eric Goldblatt. This research has been supported by the NSF
under grant number PHY-9396246 and research funds provided by The
Pennsylvania State University.

\end{document}